%% file: ms.tex
\documentclass[sigconf,nonacm,anonymous=false]{acmart}

\renewcommand\footnotetextcopyrightpermission[1]{} 
\AtBeginDocument{%
  }

\usepackage[english]{babel}
\usepackage{multirow}
\usepackage{booktabs}
\usepackage{float}
\usepackage[ruled,vlined,linesnumbered]{algorithm2e}
\usepackage{tikz}
\usetikzlibrary{shapes.arrows,positioning,arrows.meta,bending,shapes,shapes.callouts,matrix,decorations.pathreplacing,fit,backgrounds}
\usepackage{tikzsymbols}
\usepackage{graphicx}

\usepackage{color}
\definecolor{gray}{rgb}{0.5,0.5,0.5}
\usepackage{listings}
\usepackage{fancybox}

\makeatletter
\newenvironment{CenteredBox}{%
\begin{Sbox}}{
\end{Sbox}\centerline{\parbox{\wd\@Sbox}{\TheSbox}}}
\makeatother

\lstdefinelanguage{HElium}{
	keywords={true, false, catch, fun, return, null, var, if, for, in, while, do, else, input, output, plain},
	otherkeywords={<=, =>},
	keywordstyle=\color{purple}\bfseries,
	identifierstyle=\color{black},
	sensitive=false,
	comment=[l]{//},
	morecomment=[s]{/*}{*/},
	commentstyle=\color{applegreen}\ttfamily,
	stringstyle=\color{darkgray}\ttfamily,
	morestring=[b]',
	morestring=[b]",
	escapeinside={(*}{*)}
}
\usepackage{ifthen}
\newboolean{showcomments}
\setboolean{showcomments}{false}
\ifthenelse{\boolean{showcomments}}
{ \newcommand{\mynote}[3]{
		\fbox{\bfseries\sffamily\scriptsize#1}
		{\small$\blacktriangleright$\textsf{\emph{\color{#3}{#2}}}$\blacktriangleleft$}}}
{ \newcommand{\mynote}[3]{}}
\definecolor{asparagus}{rgb}{0.53, 0.66, 0.42}
\definecolor{amaranth}{rgb}{0.9, 0.17, 0.31}
\definecolor{blue}{rgb}{0.0, 0.53, 0.74}
\definecolor{applegreen}{rgb}{0.55, 0.71, 0.0}
\definecolor{amethyst}{rgb}{0.6, 0.4, 0.8}
\newcommand{\jc}[1]{\mynote{Jeronimo}{#1}{amethyst}}

\newcommand{\ls}[1]{\mynote{Lars}{#1}{amaranth}}

\usepackage{hyphenat}
\usepackage{hyperref}

\newcommand{\bnf}[1]{$\langle$#1$\rangle$}
\newcommand{\lex}[1]{\textit{#1}}

\begin{document}

\title[{HElium Compiler for Fully Homomorphic Encryption}]{HElium: A Language and Compiler for Fully Homomorphic Encryption with Support for Proxy Re-Encryption}

\author{Mirko G{\"u}nther}
\affiliation{%
  \institution{SAP SE}
  \city{Dresden}
  \country{Germany}
}
\email{mirko.guenther@sap.com}

\author{Lars Sch{\"u}tze}
\orcid{0000-0003-1422-6601}
\affiliation{%
  \institution{Technische Universit{\"a}t Dresden}
  \city{Dresden}
  \country{Germany}
}
\email{lars.schuetze@tu-dresden.de}

\author{Kilian Becher}
\affiliation{%
  \institution{Technische Universit{\"a}t Dresden}
  \city{Dresden}
  \country{Germany}
}
\email{kilian.becher@tu-dresden.de}

\author{Thorsten Strufe}
\affiliation{%
  \institution{Karlsruhe Institute of Technology}
  \city{Karlsruhe}
  \country{Germany}
}
\email{thorsten.strufe@kit.edu}

\author{Jeronimo Castrillon}
\orcid{0000-0002-5007-445X}
\affiliation{%
  \institution{Technische Universit{\"a}t Dresden}
  \city{Dresden}
  \country{Germany}
}
\email{jeronimo.castrillon@tu-dresden.de}
%
\renewcommand{\shortauthors}{M. G{\"u}nther et al.}

\begin{abstract}
Privacy-preserving analysis of confidential data can increase the value of such data and even improve peoples' lives.
Fully homomorphic encryption (FHE) can enable privacy-preserving analysis.
However, FHE adds a large amount of computational overhead and its efficient use requires a high level of expertise.
Compilers can automate certain aspects such as parameterization and circuit optimizations.
This in turn makes FHE accessible to non-cryptographers.
Yet, multi-party scenarios remain complicated and exclude many promising use cases such as analyses of large amounts of health records for medical research.
Proxy re-encryption (PRE), a technique that allows the conversion of data from multiple sources to a joint encryption key, can enable FHE for multi-party scenarios.
Today, there are no optimizing compilers for FHE with PRE capabilities.

We propose HElium, the first optimizing FHE compiler with native support for proxy re-encryption.
HElium features HEDSL, a domain-specific language (DSL) specifically designed for multi-party scenarios.
By tracking encryption keys and transforming the computation circuit during compilation, HElium minimizes the number of expensive PRE operations.
We evaluate the effectiveness of HElium's optimizations based on the real-world use case of the tumor recurrence rate, a well-known subject of medical research.
Our empirical evaluation shows that HElium substantially reduces the overhead introduced through complex PRE operations, an effect that increases for larger amounts of input data.
\end{abstract}
%


\maketitle

\input{introduction}

\input{background}
\input{scenario}

\input{architecture}

\input{evaluation}

\input{related_work}

\input{conclusion}



\bibliographystyle{acm}
\bibliography{ms}

\end{document}

%% file: introduction.tex
\section{Introduction}
\begin{figure}[t]
\centering
\includegraphics[width=.8\linewidth]{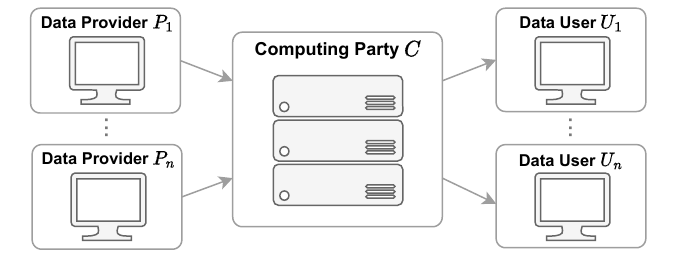}
\caption{High-Level Multi-Party Scenario with $n$ Data Providers, One Central Computing Party, and $n$ Data Users}
\label{fig:abstract_scenario}
\end{figure}

Today, healthcare institutions collect large amounts of data about their patients such as medical-treatment protocols and surgical reports. Analyses of combined data sets could increase the value of that data and provide more significant insights.
For example, with larger data sets, more accurate information about correlations between genome mutations and the properties of tumors can be identified.
This can enable researchers to develop more specific treatment methods against these tumors~\cite {Yuzawa.2016,MBraden.2014}.
However, patient data is highly confidential, and strict data-protection requirements apply~\cite{Dove.2015}. Furthermore, the pool of relevant medical data records could change over time.
Consequently, sharing and processing patients' data with research institutions for scientific studies is anything but trivial. Our work aims to solve this and similar problems by simplifying the application of privacy-preserving computation for domain experts with limited to no experience in privacy-enhancing technologies. We propose HElium, an optimizing compiler for fully homomorphic encryption. It is particularly designed for multi-party scenarios as depicted in Figure~\ref{fig:abstract_scenario}. HElium paves the way for leveraging medical data for more effective medical research without sacrificing the privacy of individuals.
Fully homomorphic encryption (FHE) schemes allow computation on encrypted data without intermediate decryption~\cite{Gentry.2009}, thus, enabling privacy-pre-\linebreak serving computation.
However, FHE is limited in terms of computations with multiple parties.
Since standard FHE requires ciphertexts to be encrypted under the same key, all parties have to agree on a common key pair before encryption. However, the generation of a joint key adds organizational overhead and requires some level of synchronicity. This might rule out many use cases in the first place, especially those that involve numerous data owners that provide confidential inputs, potentially over a longer period of time.

Approaches like multi-key FHE~\cite{LopezAlt.2012} and threshold FHE~\cite{Asharov.2012} address this problem.
Multi-key FHE enables homomorphic computation with ciphertexts encrypted under different keys and outputs a result that is encrypted under a combination of the data providers' keys.
In threshold FHE, data providers jointly run a key generation procedure that yields a common encryption key and one share of the secret decryption key per participant.
However, both multi-key and threshold FHE require data providers' active participation in the decryption procedure, which limits flexibility and convenience.
Both hinder adoption in scenarios where data providers are natural persons like patients, especially if data is added to the pool of medical data records over a longer period.

In contrast, proxy re-encryption (PRE)~\cite{Blaze.1998,Ivan.2003} enables the transformation of a ciphertext encrypted under one key into a ciphertext of the same plaintexts encrypted under a different key without intermediate decryption.
Hence, it can be used to convert ciphertexts encrypted under different keys to a common key.

FHE schemes with PRE capabilities enable computations on distributed data sets that are encrypted under different encryption keys~\cite{Polyakov.2017}. This renders them well-suited for multi-party scenarios.
Therefore, among many other problems, they allow to combine patients' health records in a privacy-preserving way.
This, in turn, allows for analyzing combined data sets to gain more significant insights, eventually leading to more effective medical research.
However, FHE and PRE operations are computationally expensive.
Therefore, they bear great potential for optimization, e.g., through parallelizing operations or rearranging their order.
This is particularly the case when individuals contribute multiple data points, for instance across different phases of treatment.

Developing efficient FHE applications with PRE is challenging and requires cryptographic expertise. 
Compilers can help developers by automating tasks like parameterization and the efficient use of re-encryption operations.
Even though a variety of compilers for FHE have been proposed in the past~\cite{Carpov.2015,Chielle.2018,Crockett.2018,Viand.2018,Dathathri.2018,Boemer.04302019,Dathathri.2020,Viand.2021,Cowan2021}, none of them natively offers optimizations for the computational overhead caused by re-encryption operations. This renders them unsuitable for scenarios with large or frequently changing participant sets.


We propose HElium, a compiler for fully homomorphic encryption with native support for proxy re-encryption.
HElium performs a plethora of optimizations such as the balancing of the computation graph as well as the minimization of the number of PRE operations.
Furthermore, we propose HEDSL, HEliums's domain-specific language (DSL) that acts as an abstraction layer for complex FHE and PRE operations and captures the different keys in multi-party scenarios.
Its syntax significantly improves the ease of use of FHE and PRE, especially for non-cryptographers.

We evaluate HElium using the real-world example of tumor recurrence rate, an active area of medical research.
We show that HElium minimizes PRE operations compared to a baseline approach for scenarios where the number of inputs $n$ is larger than the number of input keys $k$, i.e., $n/k>1$.
Most notably, for $n/k=8$, the number of PRE operations is reduced by $87.5\,\%$ compared to the baseline approach, which results in a speedup of $1.34\times$.
Consequently, our optimizations achieve high efficiency.
They ensure that the execution time overhead introduced by PRE operations quickly becomes negligible, even for relatively small $n/k$.


The structure of this paper is as follows. We first provide preliminaries such as homomorphic encryption and proxy re-encryption in Section~\ref{sec:background} followed by a detailed description of our scenario and the resulting adversary model in Section~\ref{sec:scenario}.
We give an overview of the HElium compiler architecture in Section~\ref{sec:compiler} and present the empirical evaluation of HElium in Section~\ref{sec:evaluation}. Relevant related work is discussed in Section~\ref{sec:related_work}.
We conclude the paper with a summary and approaches to future work in Section~\ref{sec:conclusion}.

%% file: background.tex
\section{Preliminaries}\label{sec:background}
This section introduces homomorphic encryption and proxy re-encryption.
While the former enables to outsource of computations to an untrusted party, i.e., a cloud service, the latter enables multi-party computations with data encrypted under different keys.

\subsection{Homomorphic Encryption}
Asymmetric cryptosystems are tuples $\mathcal{CS} = (G,E,D)$ that consist of a probabilistic key-generation algorithm $G(\cdot)$, which generates pairs of a (public) encryption key $pk$ and a (secret) decryption key $sk$, as well as a (probabilistic) encryption procedure $E(\cdot)$ and a decryption procedure $D(\cdot)$.
We further denote plaintext and ciphertext spaces by $\mathcal{M}$ and $\mathcal{C}$, respectively.

Homomorphic encryption (HE) schemes additionally provide at least one operation ``$\circ$'' on $\mathcal{C}$ that corresponds to an operation ``$\bullet$'' on $\mathcal{M}$ such that $E_{pk}(m_1) \circ E_{pk}(m_2)$ results in encryption of $m_1 \bullet m_2$ for two plaintexts $m_1,m_2$. We formalize this as
\begin{equation}\label{Eq:HomomorphicOperation}
	D_{sk}(E_{pk}(m_1) \circ E_{pk}(m_2)) = m_1 \bullet m_2,
\end{equation}
where $\bullet$ denotes Boolean or arithmetic operations like XOR, AND, addition, and multiplication as shown below for homomorphic addition and multiplication of two ciphertexts.
\begin{equation}\label{Eq:HomomorphicOperationAddition}
    D_{sk}(E_{pk}(m_1) \oplus E_{pk}(m_2)) = m_1 + m_2
\end{equation}
\begin{equation}\label{Eq:HomomorphicOperationMultiplication}
    D_{sk}(E_{pk}(m_1) \odot E_{pk}(m_2)) = m_1 \cdot m_2
\end{equation}
In addition to homomorphic operations with two encrypted operands, common HE schemes~\cite{Paillier.1999,Fan.2012,Cheon.2017} also enable operations where only one operand is provided as ciphertext whereas the second operand is a constant. These are often referred to as addition/multiplication-by-constant or ciphertext-plaintext operations. They are typically more efficient than ciphertext-ciphertext operations. We formalize them as follows.
\begin{equation}\label{Eq:HomomorphicOperationCiphertextPlaintext}
	D_{sk}(E_{pk}(m_1) \circ m_2) = m_1 \bullet m_2
\end{equation}

Homomorphic operations can be used to construct more complex functions.
Functions consisting of Boolean operations are typically interpreted as \emph{circuits} consisting of gates and wires, while arithmetic operations are often interpreted as \emph{Abstract Semantic Graphs}, also known as \emph{Term Graphs}.

There are four different types of HE schemes~\cite{Acar.2018}.
Partially homomorphic encryption (PHE) schemes offer one homomorphic operation, e.g., addition~\cite{Paillier.1999} or multiplication~\cite{Rivest.1978} of underlying plaintexts.
Somewhat homomorphic encryption (SHE) schemes enable the evaluation of two different operations, e.g., XOR and AND or addition and multiplication, but only for limited-depth circuits.
Leveled fully homomorphic encryption (LFHE) schemes offer two homomorphic operations for circuits with pre-determined bounded depths.
Fully homomorphic encryption (FHE) schemes enable the evaluation of arbitrary circuits of unlimited depth.

In 2009, Gentry presented the first FHE scheme~\cite{Gentry.2009}, which acts as a blueprint for future FHE schemes.
Public and secret key are elements that cancels out to zero during decryption~\cite{Viand.2021}.
To ensure security, a small error, referred to as \emph{noise}, is incorporated into the public key and ciphertexts.
Homomorphic operations cause the noise to grow.
As long as the noise is small enough, the plaintext can be reconstructed correctly.
If, however, the noise grows too large, it affects the underlying plaintext such that correct decryption can no longer be guaranteed.
Therefore, this construction alone is bounded to a limited number of consecutive operations.
To overcome this problem, Gentry proposed a function referred to as \emph{bootstrapping}.
This computationally complex operation reduces the noise of a ciphertext.
In the bootstrapping operation, the decryption function of the encryption scheme is translated into a circuit $\mathcal{D}$.
This bootstrapping circuit is homomorphically evaluated given the public key $pk$, the double-encrypted secret $m$, and the encrypted secret key $sk$ as follows.
\begin{equation}\label{Eq:GenericBootstrapping}
    E_{pk}(m) \leftarrow Eval(pk,\mathcal{D},E_{pk}(E_{pk}(m)),E_{pk}(sk))
\end{equation}
This yields a \emph{fresh} encryption of the plaintext, where fresh means that the resulting ciphertext contains less noise than the original ciphertext, allowing for further homomorphic operations.
Consequently, such a scheme can evaluate an arbitrary number of computation gates.


Due to the fact that homomorphic operations cause large computational overhead, FHE has been an active area of research since Gentry's seminal work.
More recent FHE schemes offer higher efficiency.
In~\cite{Brakerski.2012}, Brakerski, Gentry, and Vaikuntanathan presented a scheme that is based on the ring learning with errors (RLWE) problem~\cite{Lyubashevsky.2010}.
This and following schemes like Brakerski/Fan-Ver\-cauteren (BFV)~\cite{Brakerski.2012.Crypto,Fan.2012} and Cheon-Kim-Kim-Song (CKKS)~\cite{Cheon.2017} reduced the growth of noise during homomorphic operations.
This allows evaluation of larger circuits without bootstrapping by using the scheme in its leveled form.
\emph{Batching}, a method for packing multiple plaintexts into a single ciphertext~\cite{Smart.2014}, allows computations in a SIMD fashion, where an operation is applied to multiple plaintext elements simultaneously.
Therefore, this technique can vastly improve the efficiency of FHE-based programs.

\subsection{Proxy Re-Encryption}\label{sec:preliminaries_PRE}

Re-encryption transforms a ciphertext $c_1=E_{pk_1}(m)$ encrypted under a key $pk_1$ into a ciphertext $c_2=E_{pk_2}(m)$ of the same plaintext but encrypted under a different key $pk_2$.
Proxy re-encryption (PRE) allows an untrusted party to perform this transformation without affecting confidentiality.
PRE was first proposed by Blaze, Bleumer, and Strauss~\cite{Blaze.1998} for bidirectional settings and later by Ivan and Dodis~\cite{Ivan.2003} for unidirectional settings.

Following the notation of~\cite{Polyakov.2017}, we denote a PRE scheme as a tuple $\mathcal{PRE}=(PG,KG,ReKG,E,D,RE)$ of six procedures.
Parameter generation $PG(\cdot)$ computes a set of public parameters related to the security parameter $\lambda$.
The key generation algorithm $KG(\cdot)$ outputs a key pair $(pk, sk)$.
Re-encryption-key generation $ReKG(\cdot)$ takes a secret key $sk_i$ and a public key $pk_{j\neq i}$ and computes a re-encryption key $rk_{i \rightarrow j}$.
The re-encryption algorithm $RE(\cdot)$ transforms a ciphertext $c_i$ of $m$ encrypted under $pk_i$ into a ciphertext $c_{j}$ of $m$ such that $c_j$ encrypts $m$ under $pk_{j\neq i}$.
$E(\cdot)$ and $D(\cdot)$ denote encryption and decryption algorithms, respectively.

A generic construction to obtain a PRE scheme from an FHE scheme is described in Gentry's seminal work~\cite{Gentry.2009}. Gentry's bootstrapping construction (see Equation~\eqref{Eq:GenericBootstrapping}) implies a one-way proxy re-encryption scheme where re-encryption is conducted as follows.
\begin{equation}\label{Eq:GenericPRE}
    E_{pk_2}(m) \leftarrow Eval(pk_2,\mathcal{D},E_{pk_2}(E_{pk_1}(m)),E_{pk_2}(sk_1))
\end{equation}
Here, the re-encryption key is $rk_{1 \rightarrow 2}=E_{pk_2}(sk_1)$, i.e., the first secret key encrypted under the second public key.

%% file: scenario.tex
\section{Multi-Party Computing Scenarios}\label{sec:scenario}

This section gives an overview of the class of multi-party scenarios that our compiler enables through proxy re-encryption.

\subsection{Multi-Party Computation with PRE}\label{sec:mpc-scenario}

\begin{figure}[t]
\centering
\includegraphics[width=\linewidth]{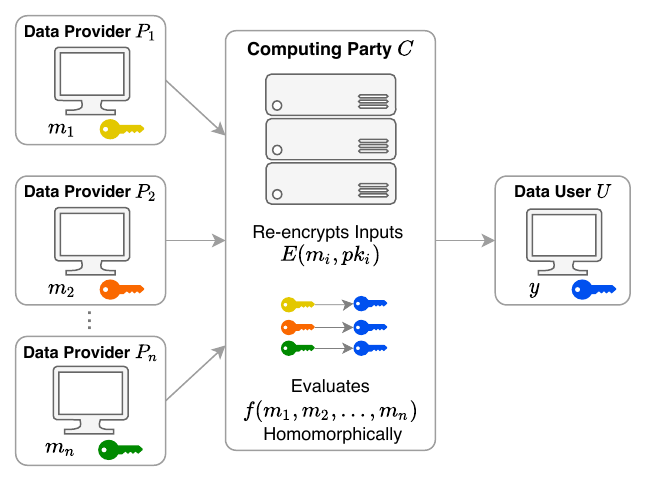}
\caption{Abstract FHE Scenario with PRE}
\label{fig:abstract-pre-scenario}
\end{figure}

Figure~\ref{fig:abstract-pre-scenario} depicts an abstract scenario of a proxy re-encryption program.
It comprises three types of participants: the data providers $P_i$ with $i\in \{1,2,\dots,n\}$, data user(s) $U$, and a computing party $C$ in between. Without loss of generality, we assume a single data user $U$ for the remainder of this paper.
The data providers have confidential data denoted as messages $m_i$.
They provide their encrypted messages to the computing party $C$, which computes the function $f(\dots,m_i,\dots)$ specified by the data user $U$. The $P_i$ could provide multiple data points $m_i$ encrypted under the same key, which could for instance be treatment details of different phases of a medical treatment.

All data providers and the data user have their key pairs to encrypt and decrypt messages.
The computing party in between enables the joint computation of the function and acts as a re-encryption proxy.
It re-encrypts the ciphertexts $E_{pk_i}(m_i)$ given the corresponding, pre-computed re-encryption keys to bring them under $U$'s public key, and homomorphically computes the function $f(\dots,m_i,\dots)$.
The result $E_{pk_U}(y)$ is forwarded to the data user.

This construction enables use cases with frequently changing data providers since new data providers only have to provide a corresponding re-encryption key.
Furthermore, it enables asynchronicity between encryption and usage of the data.
Since data is encrypted under participant-specific keys, no party can read another party's confidential data.
Neither the data user nor the target function $f(\cdot)$ with the required $m_i$'s need to be determined at the time of encryption.
The data user's identity and key pair need to be known only at the time of re-encryption-key generation.
Therefore, PRE adds a high level of flexibility concerning the participant setup.

\subsection{Threat Model}

Similar to the related work~\cite{Viand.2018,Dathathri.2019,Dathathri.2020}, we assume a semi-honest threat model~\cite{Lindell.2017} where the computing party $C$ faithfully executes the specified homomorphic evaluation but is curious about the data providers' confidential inputs. We assume that $C$ does not collude with the data user $U$.
This model is plausible in scenarios where $C$ is a server entrusted with homomorphic evaluation and an adversary $A$ has read access to $C$'s internal state without the ability to corrupt it.

Attestation of correct execution of the specified homomorphic evaluation is beyond the scope of this paper. However, following~\cite{DG17} we note that the evaluation could be computed in a trusted execution environment (TEE), like Intel SGX. This model would leverage the guarantees of correct execution that are inherent in TEEs while still relying on the strong confidentiality guarantees of homomorphic encryption.

%% file: architecture.tex
\section{HElium Compiler}\label{sec:compiler}
This section introduces the architecture of the HElium compiler. 
The HElium compiler infrastructure consists of three components: a domain-specific language (DSL) to represent multi-party FHE computations, an intermediate representation (IR), and the backend that outputs scheme-specific circuits. This section specifically highlights the use of Proxy Re-Encryption (PRE) operations (see Section~\ref{sec:preliminaries_PRE}) to facilitate multi-party scenarios.

\begin{figure}[t]
\centering
\resizebox{\columnwidth}{!}{%
\begin{tikzpicture}[
	rect/.style={rectangle,black,draw,align=center, minimum width=2cm},
	label/.style={rectangle,black, align=center, minimum width=2cm,  minimum height=0.6cm},
	outer/.style={rectangle, thick, draw=black, align=center, inner sep=0.2cm, dashed},
	]
\node[label, minimum width=0mm] (HEDSL) {HElium\\DSL\\};
\node[rect] (ProgOpt) [right=0.5cm of HEDSL] {Program\\Optimization\\(AST)};
\node[rect] (CiOpt) [right=0.5cm of ProgOpt] {Circuit\\Optimization\\(HEIR)};
\node[rect] (CryOpt) [right=0.5cm of CiOpt] {Scheme\\Optimization\\(HEIR)};

\node[rect] (BFV) [below=of CiOpt] {BFV Backend\\(C++)};
\node[label, minimum width=0mm] (Circuit) [below=of CryOpt] {FHE\\Circuit};

\node[label] (empty) [below=of ProgOpt] {};
\node[label] (empty2) [below=of ProgOpt, yshift=-0.4cm] {};

\node[label] (RT) [below=of ProgOpt, yshift=0.7cm, xshift=-0.6cm] {Runtime};
\node[label] (Comp) [above=of ProgOpt, yshift=-0.9cm, xshift=-0.6cm] {Compiler};

\node[label, minimum width=0mm] (In) [below=of HEDSL] {Secret In};
\node[label, minimum width=0mm] (Out) [below=of HEDSL, yshift=-0.4cm] {Secret Out};

\draw[-stealth] (HEDSL) -- (ProgOpt);
\draw[-stealth] (ProgOpt) -- (CiOpt);
\draw[-stealth] (CiOpt) -- (CryOpt);
\draw[-stealth] (CryOpt.south) -| (Circuit.north);
\draw[-stealth] (Circuit) -- (BFV);

\draw[-stealth] (In) -- (empty);
\draw[-stealth] (empty2) -- (Out);


\begin{pgfonlayer}{background}
	\node[outer, fit=(ProgOpt) (CiOpt) (CryOpt)] (Compiler) {};
	\node[outer, fit=(BFV) (empty) (empty2)] (Runtime) {};
\end{pgfonlayer}
\end{tikzpicture}%
}
\caption{Compiler Pipeline of HElium}
\label{fig:helium-compiler-pipeline}
\end{figure}
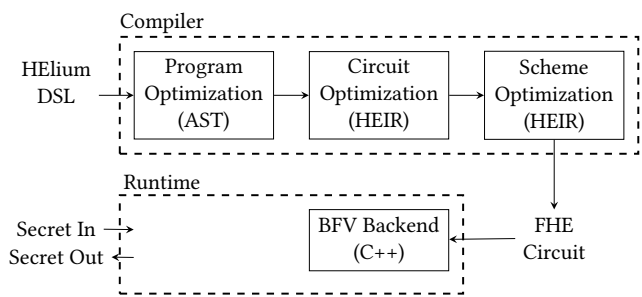

\subsection{Architecture Overview}

The compiler is designed to use staged compilation where the IR is passed through IR-specific optimizations, which are grouped into phases subsequently lowering the program representation.
Figure~\ref{fig:helium-compiler-pipeline} depicts the phases of the HElium compiler and the respective IR they work on.
The front end parses the DSL and constructs the corresponding abstract syntax tree (AST). 
The AST is de-sugared, optimized, and transformed into HElium IR (HEIR), our high-level IR representing HE circuits.
The high-level IR is subsequently transformed and optimized with attention to FHE-specific needs.
Thereafter, scheme-specific optimizations are carried out on the IR such as the automatic insertion of PRE operations.
Finally, the backend emits a circuit that can be evaluated by an FHE runtime or generates a scheme-specific IR which can be used as an input to other FHE tools.
The compiler is modular, provided as a C++ library, such that passes, phases, and backends can be added.


\subsection{HElium DSL}
\label{sec:helium-language}

This section introduces the syntax of HEDSL and highlights key features.
The HElium DSL (HEDSL) is a high-level representation of multi-party FHE programs.
The DSL hides scheme-specific details to improve usability for non-cryptographers and automates complex tasks such as scheme parameter selection.
We kept the DSL simple enough so that it can be easily embedded into mainstream programming languages.
The grammar of the DSL presented in EBNF can be seen in Figure~\ref{fig:helium_bnf}.
A HElium program consists of a list of statements, which can be, for instance, inputs, outputs, variable declarations, or function declarations.
The individual components are discussed in the following.

\begin{figure}[t]
    \centering
    \begin{tabular}{lcl}
         \bnf{Program} &:=& \bnf{Stmt}$^+$ \\
         \bnf{Stmt} &:=& \bnf{InputStmt} | \bnf{OutputStmt} \\
         &|& \bnf{VarDecl} | \bnf{VarAssignment} \\
         &|& \bnf{FuncDecl} | \bnf{ForStmt} | \bnf{ReturnStmt} \\
         \bnf{TypeExpr} &:=& plain? (int | int[ \bnf{Expr} ]) \\
         \bnf{InputStmt} &:=& input \lex{ID} : \bnf{TypeExpr} (@ \lex{ID} <= \lex{ID})? ;\\
         \bnf{OutputStmt} &:=& output \lex{ID} => \lex{ID} (@ \lex{ID} : \bnf{Expr})? ;\\
         \bnf{VarDecl} &:=& var \lex{ID} : \bnf{TypeExpr} (= \bnf{Expr})? ;\\
         \bnf{VarAssignment} &:=& \lex{ID} = \bnf{Expr} ; \\
         \bnf{ReturnStmt} &:=& return \bnf{Expr} ; \\
         \bnf{FuncDecl} &:=& fun \lex{ID} ( \bnf{ArgDecl} ) \{ \bnf{Stmt}$^+$ \} \\
         \bnf{ArgDecl} &:=& \lex{ID} | \lex{ID} , \bnf{ArgDecl} \\
         \bnf{ForStmt} &:=& for ( \lex{ID} : \bnf{Expr} ) \{ \bnf{Stmt}$^+$ \} \\
         \bnf{Expr} &:=& \bnf{Expr} ( + | - | $\ast$ ) \bnf{Expr} \\
         & | & \lex{ID} | \lex{LITERAL} | \lex{ID} (\bnf{ArgDecl}) \\
    \end{tabular}
    \caption{The grammar of the proposed HElium DSL in EBNF}
    \label{fig:helium_bnf}
\end{figure}

\subsubsection{Data Types}

HEDSL, depending on the used backend, provides primitive types for \emph{integers} and \emph{fixed-point numbers}.
It also supports \emph{fixed-size} arrays containing elements of these types.

The type system further differentiates between variables representing \emph{constants}, \emph{ciphertexts}, or \emph{plaintexts}.
Each variable is declared as ciphertext by default and captures a label of the key the ciphertext was encrypted with.
This information is later in optimization stages such as the automatic insertion of PRE operations into the circuit (see Section~\ref{sec:key-labels}).

To facilitate security by default, the usage of plaintext must be annotated explicitly to the qualifier of a variable using the \lstinline[language=HElium]{plain} keyword.
HElium uses the type information to automatically introduce more efficient ciphertext-plaintext FHE instructions when possible.

\subsubsection{Inputs and Outputs}
\label{sec:helium-language-inputs}
To support multi-party FHE, HElium programs may have multiple input and output statements each declaring the data-providing and data-consuming party, respectively.
The information is used to retrieve and assign data of different parties according to the inputs of the program.
Furthermore, data can be provided under different encryption keys.
These are captured by key labels annotated with the input statements.
The labels are used in later optimization stages such as automatic insertion of PRE operations into the circuit.

Figure~\ref{lst:dsl-example} shows the declaration of inputs. 
Similarly, outputs require a name to allow identification in programs with multiple outputs.
They are encrypted under the target key whose public key is provided to the application.
The secrets are sent back to the respective party and can be decrypted locally.

\subsubsection{Operations}

The homomorphic operations provided by the DSL work with \emph{scalar} and \emph{vector} operands.
These operations are expressed by operators as shown in Table~\ref{tab:builtin-operations}.
If one operand is a vector while the other is a scalar, the scalar will be converted into a vector first.
Operations on vectors are performed element-wise.
For rotations, the compiler provides tools to generate the respective rotation keys for participating parties.

\begin{table}[pt]
\caption{Built-in Operations of HElium}
\label{tab:builtin-operations}
\centering
\begin{tabular}{clc}
\toprule
\textbf{Symbol} & \textbf{Operation} & \textbf{Argument / Result Type} \\
\midrule
\begin{tabular}[x]{@{}c@{}}
$+$ \\
$-$ \\
$*$ \\
\end{tabular}
& 
\begin{tabular}[x]{@{}c@{}}
Addition \\
Subtraction \\
Multiplication \\
\end{tabular}
& 
\begin{tabular}[x]{@{}c@{}}
$(int,int)\rightarrow (int)$ \\
$(int[\,],int~|~int[\,])\rightarrow (int[\,])$ \\
$(int~|~int[\,],int[\,])\rightarrow (int[\,])$ \\
\end{tabular}\\
\midrule
** & Exponentiation & \begin{tabular}[x]{@{}c@{}}
$(int,int)\rightarrow (int)$\\
$(int[\,],int)\rightarrow (int[\,])$\\
\end{tabular}\\
\midrule
\textgreater{}\textgreater{} & Right rotation & \multirow{2}{*}{\begin{tabular}[x]{@{}c@{}}$(int[\,],int)\rightarrow (int[\,])$\end{tabular}} \\
\textless{}\textless{} & Left rotation  & \\
\bottomrule
\end{tabular}
\end{table}

\subsubsection{Function Definition}
To increase code reuse and encapsulation HEDSL supports the definition of functions.
Functions are defined using the \lstinline{fun} keyword followed by a function name and the function signature.
Optionally, the signature may be annotated with types. 
Figure~\ref{lst:dsl-example} depicts the definition of a function \lstinline{mul} in HEDSL.
Omitting a signature increases reuse when used in different contexts with unrelated types allowing to write functions that are polymorphic in the types of their arguments and the return type.
In such a case, the compiler creates specializations for each case where a function application applies arguments of different types not seen before.
For each specialization types of intermediate expressions are inferred according to the specification shown in Table~\ref{tab:builtin-operations}.

HElium also provides intrinsics, i.e., built-in functions, to work with built-in data types.
For example, the function \lstinline[mathescape,language=HElium]{size($\cdot$)} returns the size of a vector.
Since the size of vectors is fixed and known at compile time, HElium can directly inline the numerical value of the vector's size.

\begin{figure}[t]
\begin{CenteredBox}
\begin{lstlisting}[language=HElium]
input a1: int @Key1 <= Party1;
input b1: int @Key1 <= Party1;
input a2: int @Key2 <= Party2;
input b2: int @Key2 <= Party2;
input a3: int @Key3 <= Party3;
input b3: int @Key3 <= Party3;

fun mul(a, b) { return a * b; }

output sum: mul(a1, b1) +
            mul(a2, b2) +
            mul(a3, b3);
output quantity: b1 + b2 + b3;
\end{lstlisting}
\end{CenteredBox}
\caption{Example HElium DSL Program}
\label{lst:dsl-example}
\end{figure}

\subsubsection{Loops}

Loops help to improve readability by fostering code reuse.
HEDSL provides \lstinline[language=HElium]{for} loops, which enable the iteration over the elements of a vector.
As homomorphically encrypted programs are subject to several constraints, e.g., the absence of branching in arithmetic schemes, HElium focuses on loops with boundaries that can be determined statically.
Due to the aforementioned constraints, HElium unrolls loops during compilation.



\subsection{Intermediate Representation}
\label{sec:intermediate-representation}
A program in HElium represents operations on constants, ciphertext, or plaintext.
The HElium IR (HEIR) in-memory representation uses \emph{Term Graphs}, that is to interpret a term $t$ as a graph $g(t)$, to represent HElium programs.
In the remainder of this paper, we will use the term \emph{computation graph}.

\begin{figure}[t]
\centering
\includegraphics[scale=0.6]{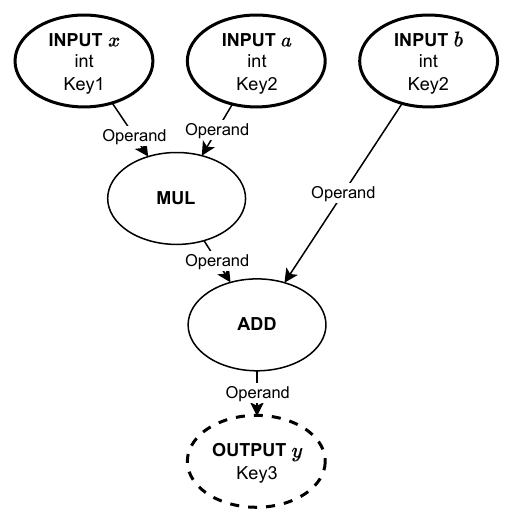}
\caption{HElium IR Graph Representation of $y = a \cdot x + b$}
\label{fig:ir-graph-example}
\end{figure}

Each term has a particular number of operands depending on the type of operation.
Unary terms take one operand as input while binary terms require two operands.
The terms \lstinline{input} and \lstinline{output} are special unary terms that not only define the dataflow in the IR but also capture the types of incoming values.
Constant values are represented by \lstinline{const} terms without an operand.
Figure~\ref{fig:ir-graph-example} shows the computation graph of an example program that calculates the function $y=a \cdot x + b$.
It has three inputs $x, a,b$ and one output $y$.

The nodes represent operations either on ciphertexts or on plaintexts.
There are special nodes that represent constant values, inputs, and outputs.
For example, the \textit{INPUT} nodes of the IR represent the inputs of the HElium program.

While the aforementioned properties are standard properties of contemporary IRs, HEDSL also captures HE-specific properties in its type system.
It is required to annotate types of (encrypted) inputs that are materialized in the computation graph.
The HElium compiler can infer types and key labels for the subsequent program.
The IR allows the HElium compiler to reason about the program and automatically injects proxy re-encryption (PRE) operations to minimize PRE operations.




\subsection{Transformations}
The parameter selection of FHE programs depends on the noise growth of ciphertexts.
As a ciphertext's noise grows with every operation performed, it is important to perform as few operations on a ciphertext as possible.
The way HE-specific components are captured in HEIR simplifies many optimizations known from existing compiler frameworks.
IRs used in contemporary compiler frameworks such as static single-assignment (SSA) forms require data-flow analyses to compute and extract information from the program not explicitly available.
In this section, we present approaches to optimizing the computation graph to improve the execution runtime.

\subsubsection{Constant Propagation and Constant Folding}
A HElium program contains constant values, either added by the developer or introduced by language constructs, e.g., loop bounds and index accesses.
While arithmetic operations with constant plaintext data are computationally less complex than ciphertext operations, pre-calculating operations with constants lowers the overall size of the computation graph.
Therefore, HElium implements constant folding, i.e., the evaluation of constant expressions at compile time.
Traditionally, this requires a reaching definition analysis to recompute \emph{use-def} and \emph{def-use} chains from the IR.
Edges between nodes in the graph directly represent use-def and use-def chains.
More precisely, our compiler evaluates if operands have use-def edges to constant values and replaces them with the constant value.
If all operands of an operation are constant the operation is statically evaluated and replaced with the result.
Due to the lack of branch instructions, this optimization only requires a single pass over the complete computation graph.

\subsubsection{Unreachable and Dead-Code Elimination}
While traditionally this requires a liveness analysis to find unused assignments, our IR uses edges between nodes to directly represent this information.
A definition without a use edge is called dead and will be removed from the graph.
This also applies to subgraphs that are not used.
Therefore, HElium has a compiler pass that removes such subgraphs iteratively.
To do so, it iterates over the drains of the computation graph and removes them, except for outputs and inputs.
Then, HElium removes the operand-usage connections to their operands.
In case their operands have no further usage, they are added to the list of drains of the computation graph.
To remove all unused nodes, this procedure is repeated until the list of drains contains only output and input nodes.
In case this list contains an input node, the compiler can notify the developer that the input is not necessary for program execution.

\subsubsection{Rebalancing}
Rebalancing can reduce the depth of the computation graph to $\mathcal{O}(log_2{n})$ given $n$ nodes in the computation graph.
In the case of multiplications, it can reduce the multiplicative depth which may have a positive impact on the parameterization and the resulting runtime performance of an HE program.
For multi-threaded evaluation, this also improves performance as sub-graphs can be evaluated in parallel.

HElium uses an approach to re-balancing similar to~\cite{chowdhary2021}.
The idea is to look for nodes that represent commutative operations like addition and multiplication.
This is implemented as a compiler pass iterating over the computation graph.
If an operation has a single usage that is of the same type of operation, both nodes are merged into a \emph{super node} of that operation.
After applying this procedure to the computation graph, all commutative chains of the same operation are transformed into super nodes with multiple operands.

The compiler then lowers the super nodes to return to a consistent IR state.
Supernodes that contain more than two operands are transformed back into a balanced computation tree.
This is applied recursively until all super nodes have been processed.

\subsection{Placing Proxy Re-Encryption Operations}
\label{sec:key-labels}
In HElium, inputs can be annotated with \emph{key labels} that declare the key input is encrypted with.
Homomorphic operations require that ciphertext operands are encrypted under the same key.
Operands with different keys must be proxy re-encrypted under the same key before they can be used for homomorphic evaluation.
The result of each homomorphic operation will be encrypted under the key of its operands.
Finally, each output must be re-encrypted under the key of the receiving party.

In a na\"ive approach, each input may be immediately proxy re-encrypted.
While this approach fulfills the requirement of operands having the same key, it is not optimal w.r.t. the amount of PRE operations inserted into the computation graph.
However, proxy re-encryption can be postponed until operands of an operation are encrypted under different keys.
This allows to reduce the amount of PRE operations, thus reducing the overall execution time.

Algorithm~\ref{alg:key-label-propagation} shows how we assign labels to nodes in the computation graph and automatically insert the least amount of PRE operations necessary.
$SUCC(v)$ denotes the nodes that outgoing edges of $v$ point to, while $PRED(v)$ denote the nodes that $v$ has incoming edges from.
The algorithm takes as input all nodes that represent the input to the computation graph.
These nodes are already labeled with the key of the ciphertexts.
A forward traversal computes whether the keys of operands differ.
If they do not, the operand's key is propagated to that node.
Otherwise, the key of the receiving party is assigned and a PRE operation for each operand is inserted into the graph.

\begin{algorithm}[tp]
	\SetKwInput{KwData}{Input}
    \KwData{The computation graph with annotated nodes $V_{in}$, and the output node $V_{out}$.}
    \KwResult{The computation graph with PRE operations.}
    \SetKwFunction{SUCC}{SUCC}
    \SetKwFunction{PRED}{PRED}
    \SetKwFunction{KEY}{KEY}
    \SetKwFunction{PRE}{PRE}
    nodes $\gets$ \SUCC($V_{in})$\;
    \ForEach{n $\in$ nodes}{
        nodes $\gets$ nodes $\setminus$ \{n\} $\cup$ \SUCC(n)\;
        keys $\gets$ $\varnothing$\;
        \ForEach{operand $\in$ \PRED(n)}{
            keys $\gets$ keys $\cup$ \KEY(operand)\;
        }
        \If{$\lvert$keys$\rvert$ = 1}{
            \KEY(n) $\gets$ \KEY(operand)\;
        }
        \If{$\lvert$keys$\rvert$ > 1}{
            \KEY(n) $\gets$ \KEY($V_{out}$)\;
            \ForEach{operand $\in$ \PRED(n)}{
                \PRE(operand, n)\;
            }
        }
    }
    \caption{PRE Operation Insertion Algorithm}
    \label{alg:key-label-propagation}
\end{algorithm}

\subsection{Backend}\label{sec:backends}
The backend of HElium translates the IR to a low-level HE instruction set which is executable by the targeted FHE scheme implementation.
During that lowering, HElium directly maps to or encodes operations of the IR in the instructions of the given scheme.
The backend collects metrics of the generated computation graph such as the minimal necessary plaintext size or the multiplicative depth.
These metrics are used to suggest suitable encryption parameters.
However, the user is not forced to use the suggested parameters.
In scenarios where the data is already encrypted before compilation, developers can provide their parameters.

HElium and its IR are designed to be scheme-agnostic.
The staged compilation and lowering allow for the extension of HElium by registering new backends.
Each backend must register capabilities and its HE instruction set.
HElium automatically assists with scheme decisions based on the capabilities of the backends.
For example, we implemented backends for the TFHE~\cite{Chillotti.2016} and CKKS~\cite{Chen.2017} schemes to integrate support for efficient Boolean operations and fixed-point arithmetic.
However, their discussion is beyond the scope of this work.

%% file: evaluation.tex
\section{Evaluation}\label{sec:evaluation}
This section presents the evaluation of the effectiveness of the optimizations provided by the HElium compiler.
We demonstrate their effectiveness in a use-case taken from the aforementioned health-care scenario.

\subsection{Use-Case: Tumor Recurrence Rate}
\label{sec:scenario-cancer}
Hospitals, universities, and private healthcare companies collect a large amount of data about their patients.
This includes, for example, protocols from medical treatments, drug applications, and surgical reports.
With the introduction of the electronic medical file, more information will be digitalized~\cite{Cowie.2017}.
This information about patients is highly confidential and high data protection requirements apply~\cite{Dove.2015}.
Therefore, sharing and aggregation of patient data is very complicated due to regulatory reasons.
FHE in conjunction with PRE can enable researchers to combine encrypted patient-data sets with other researchers without sacrificing patients' privacy.
It allows performing analyses of the shared data sets without disclosing individual patient records.

For example, records of genome mutations can be compared with histology analysis or surgical reports of tumor patients to obtain information about correlations between specific genome mutations and the properties of the tumor~\cite{MBraden.2014,Yuzawa.2016}.
That enables researchers to develop more specific treatment methods for tumors.
One measure that is analyzed in the context of cancer research is the recurrence rate of variants of tumors. 
It represents the percentage of patients for whom particular cancer reappears.
This measure indicates the medical treatment of the cancer patient.
The particular tumor type can often only be determined by surgery and the following histology analysis.
A non-invasive determination of the tumor type and its properties can allow treatment of the tumor without surgery.
Therefore, finding correlations between genome mutations and tumor properties is an active field of research.

The tumor recurrence rate $r$ of patients is calculated as an average of the recurrence of a tumor in relation to the absolute number of patients with the same tumor variant. 
Namely,
\begin{equation}
    r = \frac{\text{Number of Tumor Patients with Recurrence}}{\text{Number of Tumor Patients}}
    \label{eq:cancer-recurrence-rate1}
\end{equation}
In order to find correlations between genome mutations and the tumor recurrence rate, $r$ can be calculated with the presence of different mutations.
The presence of mutations for the $i$-th patient $P_i$ can be efficiently encoded as bits of a bit vector denoted by  $\mathbf{b}_i$. 
Each element $b_j$ of $\mathbf{b}_i$ represents the presence or absence of a mutation $j$.
Similarly, the tumor recurrence of a patient $i$ can be encoded as a single bit $a_i$.
The recurrence rate $r$ in relation to the presence of mutations is defined as shown in Equation~\eqref{eq:cancer-recurrence-rate}. 
Each element of the result vector $\mathbf{r}$ represents the recurrence rate in relation to the $j$-th mutation.
This measure can help to find correlations between the presence of mutations and the recurrence of tumors.
\begin{equation}
    \mathbf{r} = \frac{\sum_{i=0}^n a_i \cdot\mathbf{b_i}}{\sum_{i=0}^n{\mathbf{b_i}}}
    \label{eq:cancer-recurrence-rate}
\end{equation}
 
Figure~\ref{lst:use-case2-sourcecode}, shows the HElium code of an example implementation of the presented use case.
Each patient data set consists of two values $a_i$ and $b_i$ that are provided under a unique key $Key_i$.
\begin{figure}[t]
\begin{CenteredBox}
\begin{lstlisting}[language=HElium]
input a0: int       @Key0 <= Party0;
input b0: int[1000] @Key0 <= Party0;
input a1: int       @Key1 <= Party1;
input b1: int[1000] @Key1 <= Party1;
input a2: int       @Key2 <= Party2;
input b2: int[1000] @Key2 <= Party2;
input a3: int       @Key3 <= Party3;
input b3: int[1000] @Key3 <= Party3;

output R => Party_Out @Key_Out:
    a0*b0 + a1*b1 + a2*b2 + a3*b3;
output n => Party_Out @Key_Out:
    b0 + b1 + b2 + b3;
\end{lstlisting}
\end{CenteredBox}
\caption{HElium Program: The Recurrence Rate of Tumors. For Simplicity $n = 4$.}
\label{lst:use-case2-sourcecode}
\end{figure}
In lines 1 to 8, the inputs are defined for four data sets of four different participants.
The outputs and the function are declared in lines 10 and 11.
The division of two unknown ciphertexts is an operation with high computational complexity.
Therefore, the example program computes only the two sums with FHE.
The division is performed afterwards on client side as a plaintext operation.

\jc{The use case is a bit disappointing. It sounds very cool from the name, but it is, in the end, a very simple computation. To bring the point stronglier, maybe it would be good to show what a non-expert would have to write in C++ or so to achieve the same functionality.
If possible, also talk about changes that would have to be done to change the backend in the hand-written version}
\jc{A real use case is very good, but given its simplicity, I would complement the evaluation with a set of similary simple but synthetic computations (std-dev, max-min, mac, ...)} 
\ls{Maybe I should sit down and write such a program with PALISADE C++ directly and then compare ours to that?}

\subsection{Evaluation Setup}
All experiments were conducted on a dedicated server with 8 Intel(R) Core(TM) i7-9700T CPU cores and 32 GB of RAM.
Each experiment is performed $100$ times.
The runtimes of compilation and execution are measured, and the average, as well as the standard deviation, are calculated. After each run, the results are decrypted and compared with the expected result of the corresponding plaintext function to verify correctness of the compiled program.
The experiments are conducted for set sizes $n \in \{128,256,512,768,1024\}$. 
This relates to typical data-set sizes of medical research~\cite{Yuzawa.2016}. Often, only a few hundred patient records are available at a research institution depending on the type of tumor.
The set sizes also allow to showcase the impact of HElium's optimizations such as the minimization of PRE operations and the balancing of computation graphs. 
Data-set sizes of power of two are well-suited to demonstrate the effectiveness of rebalancing. 
However, our compiler is not limited to those and supports arbitrary set sizes.

\subsection{Efficiency of the PRE Integration}
This section discusses how efficiently the compiler chooses to insert PRE operations.
The compiler aims to insert only as few PRE operations as necessary.
A naive approach of inserting PRE operations is to immediately proxy re-encrypt all inputs to a common key.
We refer to the number of introduced PRE operations in this naive approach as $p_{naive}$.
It is determined by the number of inputs $i$ of the program, i.e., $p_{naive} = i$.

From a theoretical perspective, the minimal number of PRE operations $p_{min}$ is determined by the keys of the inputs and outputs of the program, as shown in Equation~\eqref{eq:number-pre-min}. 
$K_I$ and $K_O$ denote the set of input and output keys, respectively.
In the best case, only one PRE operation is needed for all keys that are members of $K_I$ but not of $K_O$, i.e., the difference of $K_I$ and $K_O$.
\begin{equation}
    p_{min} = \lvert K_I \setminus K_O \rvert
    \label{eq:number-pre-min}
\end{equation}
However, $p_{min}$ does not take the structure of a program into account.
Often, this theoretic minimum cannot be reached.
HElium aims to achieve a number of PRE operations $p$ close to $p_{min} \leq p \ll p_{naive}$.

We analyze the efficiency of HEliums approach to optimizing the insertion of PRE operations by measuring the number of inserted PRE operations and the resulting execution time of the evaluation of the HE circuit.
Furthermore, the ratio between data sets $n$ and keys $k$, denoted as $n/k$, has a significant impact on the optimization possibilities.
To explore the effects of $n/k$ on $p$ and the evaluation run-time, we conduct measurements for $n/k \in \{1,2,4,8\}$.

\begin{figure}
\centering
\includegraphics[width=\linewidth,keepaspectratio]{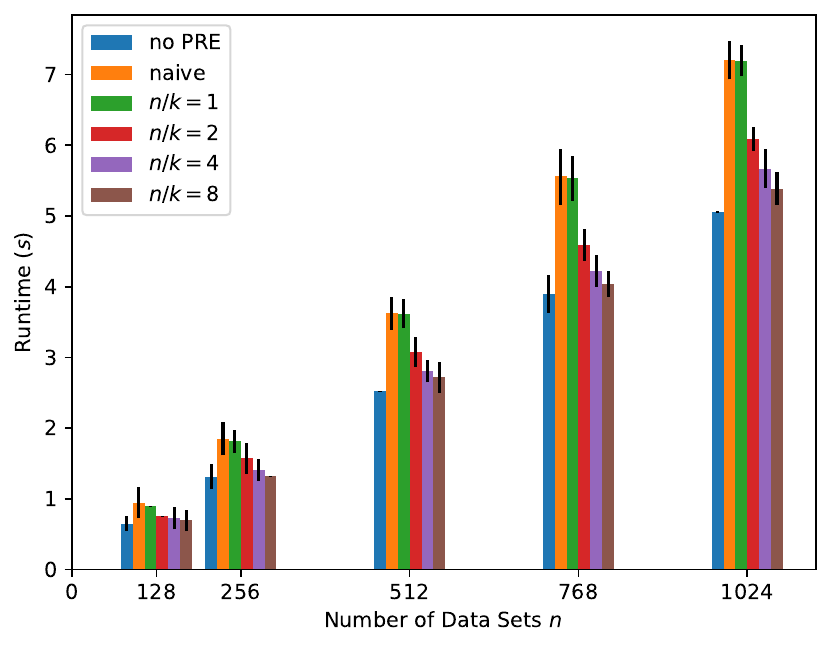}
\caption{Evaluation Run-Time}
\label{fig:evaluation-runtime}
\end{figure}

Figure~\ref{fig:evaluation-runtime} depicts the measured execution times of data-set sizes $n$ and ratios $n/k$.
Furthermore, the computation of each data-set size $n$ without PRE and the naive approach is shown to emphasize the impact of our proposed optimizations.
The larger $n/k$, the higher is the potential for optimization, hence, the smaller is the overhead introduced by PRE.
The four right-most bars for each $n$ show that the overhead quickly becomes negligible, even for relatively small $n/k$.

For $n/k=1$, each data set is provided encrypted under its own key.
In that case each input needs to be proxy re-encrypted in both approaches, i.e., $p=p_{naive}$ and their evaluation run-time is on par with each other. 
With increasing $n/k$ ratio, the execution time decreases due to a lower number of inserted PRE operations.
For example, in $n/k=2$, $p$ is reduced by $50\,\%$ compared to $p_{naive}$.
In the case of $n/k=8$, $p$ is reduced by $87.5\,\%$ which results in a speedup of $1.34\times$.
Most notably, optimizing PRE operations reduces the run-time overhead from $42.4\,\%$, for the naive variant, to $6.5\,\%$ for a ratio of $n/k=8$.

In summary, our analysis shows that the HElium compiler effectively minimizes PRE operations compared to the naive approach for scenarios $n/k>1$.
Consequently, due to our optimizations, the execution time overhead introduced by PRE operations quickly becomes negligible, even for relatively small $n/k$.

%% file: related_work.tex
\section{Related Work}\label{sec:related_work}
There exist a variety of (open-source) software libraries that implement one or multiple FHE schemes.
Compared to domain-specific compilers, these libraries often miss domain-specific analyses and optimizations.
Often being implemented using low-level programming languages like C++ increases the cognitive load to write efficient and secure programs.
However, their APIs and data structures often capture domain-specific aspects which are seen in both library implementations and external DSLs.

\subsection{Software Libraries for Fully Homomorphic Encryption}
One of the first efficient software libraries is HElib~\cite{halevi2014,Halevi2020}, providing implementations of the BGV~\cite{Brakerski.2012} and CKKS~\cite{Cheon.2017} schemes.
It offers bootstrapping for BGV and focuses on effective use of ciphertext packing and so-called ``GHS optimizations''~\cite{gentry2012}.

The Simple Encrypted Arithmetic Library (SEAL)~\cite{Chen2017} software library provides the BGV, CKKS, and BFV~\cite{Brakerski.2012.Crypto,Fan.2012} schemes and also supports bootstrapping.
It supports various platforms such as Android, iOS, and embedded systems.
SEAL provides abstractions for batching to implement more efficient homomorphic computations.
While offering an easy-to-use API, the difference between efficient and inefficient implementations can be several orders of magnitude.

PALISADE~\cite{Halevi.2019} supports the BGV, BFV, CKKS, and FHEW~\cite{Ducas.2015} schemes and a more secure variant of the TFHE scheme, including bootstrapping.
Another important feature is the support for proxy re-encryption (PRE).
However, developers need to remember the keys to be used for PRE operations.
In our approach, the target key to re-encrypt is implicit from the input program.

Lattigo~\cite{lattigo} is a software library written in Go implementing the CKKS and the BFV schemes.
Similar to PALISADE, Lattigo provides threshold variants of the implemented schemes.

The TFHE-rs library~\cite{TFHE-rs} provides a Rust implementation of the TFHE scheme~\cite{Chillotti2018}.
TFHE is optimized for Boolean and integer arithmetics and fast bootstrapping. 
Among others, TFHE-rs offers programmable bootstrapping~\cite{chillotti2021programmable}. 
This technique allows the application of unary functions during bootstrapping via lookup tables. 
It is a technology to enable efficient computation of non-linear functions like square roots or activation functions like ReLU.
The ability to compute non-linear functions is a big advantage for machine learning use cases.

\subsection{Compilers for Fully Homomorphic Encryption}
To the best of our knowledge, HElium is the first compiler that optimizes proxy re-encryption. Existing compilers often simply assume that data providers agree on a common key before homomorphic evaluation. However, agreeing on a joint key among multiple parties adds organizational overhead and requires some level of synchronicity. Especially the latter might already exclude many use cases where data is added over a longer period and a joint key agreement might be hard to perform. 
Thus, in this section, we focus on approaches to improving the generated circuits that are going to be homomorphically evaluated.
We do not compare to work on exploiting parallelism on hardware platforms such as GPUs and FPGAs to improve the performance of homomorphic computations as this is beyond the focus of this work.
See \cite{Viand.2021} for a more detailed comparison of FHE compilers in general.

Some approaches use DSLs to capture FHE programs as computations over simple expressions, generate related circuits, and link them with FHE backends.
This approach is also used in Alchemy~\cite{Crockett.2018}, a language and compiler integrated into Haskell.
The integrated compiler automatically inserts maintenance operations into the program and selects suitable encryption parameters.
It uses its implementation of the BFV scheme. 
Encrypt-Everything-Everywhere (E3)~\cite{Chielle.2018} uses template meta-programming to generate circuits from C++ code.
For each supported backend it provides specializations of the templated classes.
It also provides a collection of auxiliary tools to assemble an executable from configuration files. 

Another class of approaches has a strong focus on FHE for machine learning applications and tensor operations.
The Compiler and Runtime for Homomorphic Evaluation of Tensor Programs (CHET)~\cite{Dathathri.2019}  comes with a high-level language for machine-learning operations.
It focuses on optimizations for matrix operations using efficient batching and optimizes scaling factors automatically, given a set of test inputs.
EVA~\cite{Dathathri.2020,chowdhary2021} is a low-level compiler for vector-arithmetic operations.
It extends the idea of CHET and is a C++ library with a Python interface that targets the CKKS implementation of SEAL.
EVA inserts maintenance operations into the computation graph using a cost model and automatically selects encryption parameters.
HECATE~\cite{lee2022} provides a DSL and Intermediate Representation (IR) to optimize scale management of circuits using the CKKS scheme.
It presents a flexible approach to rescaling that explores multiple different rescaling approaches as well as a preconstructed estimation of the latency of each FHE operation at different levels and polynomial modulus $N$.
This estimator is used to estimate the performance of a HECATE IR program during optimization.
nGraph-HE~\cite{Boemer.04302019} is a homomorphic encryption backend to the Intel nGraph compiler for machine learning.
It focuses on the inference of machine learning models over encrypted data.
nGraph-HE translates Tensorflow computations into FHE circuits for BFV or CKKS.
It applies optimizations on the computation graph and supports SIMD-packing, i.e., the efficient packing of multi-dimensional tensors into batched ciphertexts.
HECO~\cite{viand2023} provides an end-to-end architecture from a high-level DSL down to hardware-accelerated FHE circuits.
The compiler infrastructure is based on MLIR~\cite{lattner2020mlir} and optimizations focus on the automatic generation of batched HE instructions.

A different approach is taken by Porcupine~\cite{Cowan2021}.
It employs program synthesis to automatically generate vectorized homomorphic encryption programs from a high-level specification of a kernel and a sketch of that kernel.
A sketch consists of operands which may be defined as holes and possible arithmetic instructions used to implement the specification.
Holes in the sketch define the space that is searched.
To reduce the search space, rotation is not an instruction living on its own but is considered as input to arithmetic instructions in the sketch.
An SMT solver verifies that a solution of a sketch fulfills the specification.

%% file: conclusion.tex
\section{Conclusion}\label{sec:conclusion}

Efficient privacy-preserving analysis of confidential data, e.g., for medical research, is a pressing problem of our time.
Fully homomorphic encryption (FHE) schemes allow computations over encrypted data.
FHE has proven to be a valuable means to leverage the value of data without sacrificing the privacy of individuals.
However, FHE operations are computationally expensive and optimizations require a high level of expertise.
Compilers for FHE can automate many optimization aspects and improve usability of FHE, especially for non-cryptographers.
However, FHE operations require data to be encrypted under the same key.
Yet, many real-world use cases imply multi-party scenarios where data is encrypted under different keys, thus, excluding standard FHE for numerous use cases.

FHE schemes with proxy re-encryption (PRE) capabilities allow the conversion of ciphertexts encrypted under different keys into ciphertexts encrypted under a common key, thus, filling the above-mentioned gap.
The automated optimization of FHE programs with PRE operations, however, remains yet to be solved.
We target this open question and propose HElium, the first compiler for fully homomorphic encryption with native proxy re-encryption support.
HEliums's domain-specific language acts as an abstraction layer for complex FHE and PRE operations and, therefore, significantly improves the ease of use of FHE and PRE, especially for non-cryptographers.
Among a variety of optimizations, our HElium compiler minimizes of the number of PRE operations and conducts balancing of the computation graph.
This substantially improves the ease of use of fully homomorphic encryption in multi-party scenarios, especially for non-cryptographers.

Based on the real-world example of tumor recurrence rate, an active area of medical research, we show that HElium minimizes PRE operations compared to a baseline approach for scenarios where the number of inputs $n$ is larger than the number of input keys $k$, i.e., $n/k>1$.
Most notably, for $n/k=8$, the number of PRE operations is reduced by $87.5\,\%$ compared to the baseline approach, which results in a speedup of $1.34\times$.
Consequently, our optimizations ensure that the execution time overhead introduced by PRE operations quickly becomes negligible, even for relatively small $n/k$.

In the future we plan to further increase performance by extending the runtime of HElium to evaluate homomorphic circuits in parallel.
We see a promising future research direction in adding a cost-model based optimization taking both performance and noise accumulation into account.
The compiler cannot reason about these independently, as performance cost often depends on noise growth because of the underlying mathematical homomorphic encryption instruction implementations.